\documentclass[12pt]{iopart}
\usepackage{iopams}
\usepackage{graphicx}

\newcommand{\dd}{{\mathrm{d}}}   
\newcommand{\ee}{{\mathrm{e}}}   

\begin{document}

\title[Regular and quasi black hole solutions for charged dust]{Regular
and quasi black hole solutions
for spherically symmetric charged dust distributions
in the Einstein--Maxwell theory}

\author{ {Dubravko Horvat}, {Sa{\v{s}}a Iliji{\'c}}
    and {Zoran Naran{\v{c}}i{\'c}} }

\address{Department of Physics,
   Faculty of Electrical Engineering and Computing,
   University of Zagreb, Unska~3, HR-10\,000 Zagreb, Croatia}

\ead{dubravko.horvat@fer.hr,sasa.ilijic@fer.hr}

\date{\today}

\begin{abstract}
Static spherically symmetric distributions
of electrically counterpoised dust (ECD) are used to construct solutions
to Einstein-Maxwell equations in Majumdar--Papapetrou formalism.
Unexpected bifurcating behaviour of solutions
with regard to source strength is found for localized,
as well as for the delta-function ECD distributions.
Unified treatment of general ECD distributions is accomplished
and it is shown that for certain source strengths
one class of regular solutions approaches Minkowski spacetime,
while the other comes arbitrarily close to black hole solutions.
\end{abstract}


\maketitle


\section{Introduction \label{sec:intro}}

Static bodies resulting from combined gravitational attraction
and electrical (electrostatic) repulsion
have been playing a central role in many investigations,
eg.\ \cite{BoWick75,BoWick72,Bonnor98} and references therein,
connected with general relativistic treatment of gravitational contraction.
As a part of Einstein equation,
one has to specify energy-momentum tensor
which, for a perfect fluid, has the form
  \begin{equation}
  T_{\mu\nu} = (\rho+p) \, u_{\mu}u_{\nu} +p \, g_{\mu\nu},
  \label{eq:T1}
  \end{equation}
where $\rho$, $p$, $u^{\mu}$ and $g_{\mu\nu}$
are mass density, pressure, four-velocity and metric
(in geometrized units: $G=1=c$).
In addition, one has to specify the equation of state
relating the pressure to the density $p=p(\rho)$.
According to the Oppenheimer-Snyder scenario \cite{OppenSny39},
in a static spherically symmetric perfect fluid ball
with the above energy momentum tensor,
after stationary phase and at the end of nuclear burning,
the central pressure is no longer able to counterbalance
the gravitational attraction and the ball starts to collapse.
By setting $p=0$ the energy momentum tensor assumes the form
characteristic for dust ball $T_{\mu\nu}=\rho u_{\mu} u_{\nu}$.
Eventually, as the ball shrinks
one can expect formation of a black hole.

The Majumdar--Papapetrou formalism \cite{Maj47,Papa47,Papa54,Varela03}
can be applied to study properties of spacetimes
generated by $T_{\mu\nu}=\rho u_{\mu} u_{\nu}$
where $\rho$ represents static electrically counterpoised dust
(or extremal charged dust, ECD),
i.e.\ pressureless matter in equilibrium under its own
gravitational attraction and electrical repulsion
\cite{BoWick75,BoWick72,Bonnor98}.
Special attention has been paid to ECD distributions
leading to spacetimes that are regular everywhere,
but with exterior arbitrarily close to that
of an extremal Reissner--Nordstr{\o}m (ERN) black hole
\cite{Bonnor99,LeWein04,KZL04}.

This paper considers the coupled Einstein-Maxwell field equations
in the Majumdar--Papapetrou approach.
With the energy-momentum tensor given by (\ref{eq:T1}) with $p=0$,
our study will concentrate on diverse (analytic)
spherically symmetric forms of ECD distributions
and corresponding solutions will be found.
The theoretical framework is presented in section~\ref{sec:rev}.
In section~\ref{sec:bif}, we present
the static solutions to Einstein-Maxwell equations
for diverse ECD distributions obtained numerically,
and in section~\ref{sec:delta} for the $\delta$-shell distribution,
all of which show interesting bifurcating properties.
In section~\ref{sec:uni},
we discuss the bifurcating behaviour and
show that diverse distributions can be treated on equal footing.
Similar behaviour found in gauge theories
and Einstein--Yang--Mills systems is discussed,
and we give some details on numerical methods
for finding bifurcating solutions.
In section~\ref{sec:qbh}, we show that with adjusting of the source strength
and appropriate rescaling of the solutions,
the spacetimes with exteriors arbitrarily close to the ERN case
can be obtained.
Conclusions are given in section~\ref{sec:concl},
and extension of the present work is proposed.


\section{Theoretical framework \label{sec:rev}}

One usually starts by defining
a static spherically symmetric line element of the form
  \begin{equation}
  \dd s^2 = - B(r) \, \dd t^2
            + A(r) \, \dd r^2
            + r^2  \, \dd \omega^2 ,
  \label{eq:ds2:AB}
  \end{equation}
where $\dd\omega^2 = \dd\vartheta^2 + \sin^2\vartheta \, \dd\varphi^2$.
For an asymptotically flat spacetime, one requires
  \begin{equation}
  A(r) \big|_{r\to\infty} = B(r) \big|_{r\to\infty} = 1 ,
  \label{eq:AB:inf}
  \end{equation}
and for a nonsingular spacetime $A(r=0)=1$.
Papapetrou has shown \cite{Papa47,Papa54}
that the line element can be written as
  \begin{equation}
  \dd s^2 = - \ee^{\chi}    \, \dd t^2
            + \ee^{\varphi} \, ( \dd X^2 + \dd Y^2 + \dd Z^2 )
  \label{eq:ds2:papa}
  \end{equation}
and that it is possible to connect functions $\chi$ and $\varphi$
in such a way as to satisfy (as one possibility)
$\dd\chi/\dd\varphi=-1$ or $\chi(\varphi)=-\varphi$.
Together with Majumdar's assumption \cite{Maj47}
about connection between $g_{tt}$ (or $g_{00}$) metric component
and a scalar component of the electromagnetic potential $A^{\mu}$
comprising the electromagnetic energy-momentum tensor
$T_{\mu\nu}^{\mathrm{(em)}}$,
the line element $\dd s^2$ in the Majumdar--Papapetrou form
can be written in the harmonic coordinates $(t,\mathbf{X})$ as
  \begin{equation}
  \dd s^2 = - \ee^{-\varphi} \, \dd t^2
            + \ee^{\varphi}  \, \left( \dd X^2 + \dd Y^2 + \dd Z^2 \right)
  \label{eq:ds2:maj}
  \end{equation}
(in the original notation of \cite{Papa54}).
We will assume the spherical symmetry
as a general symmetry requirement for our problem.
The essential ingredient to the line element (\ref{eq:ds2:maj}),
as shown by Majumdar,
are additional assumptions which should be made about a specific
form for the energy-momentum tensor $T_{\mu\nu}$
that enters the Einstein equations
  \begin{equation}
  R_{\mu\nu} - \frac{1}{2} g_{\mu\nu}R = 8\pi \, T_{\mu\nu}
  \label{eq:ein}
  \end{equation}
(in geometrized units).
The component of $T^{\mu\nu}$ due to electromagnetic fields is given by
  \begin{equation}
  T^{\mathrm{(em)}}_{\mu\nu} =
    F_{\mu}^{\hphantom{\mu}\sigma} F_{\nu\sigma}
    - \frac{1}{4} \, g_{\mu\nu} \, F^{\rho\sigma} F_{\rho\sigma},
  \label{eq:Tem}
  \end{equation}
where $F_{\mu\nu} = A_{\nu,\mu} - A_{\mu,\nu}$
is the antisymmetric electromagnetic field tensor,
comma denoting the ordinary derivative
to differ from semicolon denoting the covariant derivative.
It satisfies the empty space equation $F^{\mu\nu}_{\ \ ;\nu}=0$.

A more general situation differing from the one described above corresponds to
  \begin{equation}
  T_{\mu\nu} = T_{\mu\nu}^{\mathrm{(em)}} + T_{\mu\nu}^{\mathrm{(m)}},
  \label{eq:T2}
  \end{equation}
where the superscript $(\mathrm{m})$ denotes matter.
The absence of $T_{\mu\nu}^{\mathrm{(m)}}$
leads to Reissner--Nordstr{\o}m solutions
with $B(r)=A(r)^{-1}=(1-2m/r+q^2/r^2)$ for $A^0=e/r$.
It is shown in Ref.~\cite{Bonnor60} that for charged sphere of radius $a$,
for $r$ large enough, $m$ could be set equal to zero,
whereas for the interior solution, i.e.~$r<a$,
one is not permitted to put $m=0$.
It follows that electric charge contributes
to the gravitational mass of the system.
In addition, mass as given in $B(r)=A(r)^{-1}$ above
satisfies a positivity condition $m>0$.
Citing Ref.~\cite{Bonnor60} `a charged sphere must have a positive mass'.

In the case of the complete $T^{\mu\nu}$ given in (\ref{eq:T2})
one assumes also a perfect fluid of classical hydrodynamics
for which, in general, energy-momentum tensor is given by (\ref{eq:T1}).
With $p=0$, the energy momentum tensor (\ref{eq:T1}) reduces to
$T^{\mathrm{(m)}}_{\mu\nu}=\rho \, u_{\mu}u_{\nu}$,
where $\rho$ is now an invariant dust density
and $u_{\mu}$ is the four-velocity
which will be assumed to be in non-moving (co-moving) coordinate system
so that
  \begin{equation}
  u^{\mu} = \frac{\dd X^{\mu}}{\dd s}
    \longrightarrow u^{\mu}=\delta^{\mu}_0 \,(-g_{00})^{-1/2} .
  \end{equation}
The electromagnetic part of $T_{\mu\nu}$
is given, as before, with (\ref{eq:Tem}).
The four-current density
of the the charge/matter/dust distribution is given by
  \begin{equation}
  J^{\mu}=\rho_e u^{\mu}.
  \label{eq:Jmu}
  \end{equation}
At this point, the crucial assumption is made
by setting the electric charge density $\rho_e$
equal to the matter density $\rho$,
or somewhat more general $\rho_e=\pm \rho$.
Now the metric (\ref{eq:ds2:maj}) can be written in the form
  \begin{equation}
  \dd s^2 = - U^{-2} \, \dd t^2
            + U^2    \, \left( \dd X^2 + \dd Y^2 + \dd Z^2 \right) \; .
  \label{eq:ds2:U:XYZ}
  \end{equation}
The Einstein field equations
incorporating the above introduced ingredients
are given by
  \begin{eqnarray}
  G_{\mu\nu} & = & 8\pi \,
  ( T_{\mu\nu}^{\mathrm{(em)}} +T_{\mu\nu}^{\mathrm{(m)}} ) \nonumber \\
  & = & 8\pi \, \left[
      \frac{1}{4\pi} \left(
          F_{\mu}^{\hphantom{\mu}\sigma} F_{\nu\sigma}
          - \frac{1}{4} \, g_{\mu\nu} \, F^{\rho\sigma} F_{\rho\sigma}
      \right) + \rho_e u_{\mu} u_{\nu}
  \right],
  \end{eqnarray}
and by using the Majumdar (or Majumdar--Papapetrou) assertion,
the generalized (nonlinear) Poisson equation (with $\rho=\rho_e$)
  \begin{equation}
  F^{\mu\nu}_{\hphantom{\mu\nu};\nu}
  = 4\pi \, J^{\mu} = 4\pi \, \rho u^{\mu}
  \end{equation}
assumes a simple form
  \begin{equation}
  \nabla^2 U = - 4\pi \, \rho \, U^3 .
  \label{eq:LapU}
  \end{equation}
Here $U$ and $\rho$ are
functions of the co-moving coordinates $(X,Y,Z)$
determined from the zeroth component
of the electromagnetic potential $A^{\mu}$, i.e.
  \begin{equation}
  A^{\mu} = \delta^{\mu}_0 \, \phi
  \qquad \mathrm{and} \qquad
  U = \frac{1}{1-\phi} \; .
  \end{equation}
Confining our attention to the spherically symmetric case,
the Majumdar--Papapetrou metric can be written as
  \begin{equation}
  \dd s^2 = - U^{-2}(R) \, \dd t^2 
    + U^2(R) \, \left( \dd R^2+ R^2 \, \dd \Omega^2 \right) .
  \label{eq:ds2:U:R}
  \end{equation}
The metric function $U$ is a function of $R$ only,
so (\ref{eq:LapU}) reduces to
  \begin{equation}
  \nabla^2 U(R) = R^{-2} \left( R^2 \, U'(R) \right)'
               = - 4\pi \, \rho(R) \, U^3(R),
  \label{eq:LapU:R}
  \end{equation}
where the prime denotes the differentiation with respect to $R$.
The line element (\ref{eq:ds2:U:R})
can be expressed in the standard form (\ref{eq:ds2:AB})
through the coordinate transformation
  \begin{equation}
  r=R\,U(R) ,
  \label{eq:r=UR}
  \end{equation}
with the following relations among the profile functions:
  \begin{equation}
  B(r)=U^{-2}(R)
  \label{eq:B:U}
  \end{equation}
and
  \begin{equation}
  \frac{1}{\sqrt{A(r)}} = 1 + \frac{R}{U(R)} \, \frac{\dd U(R)}{\dd R} .
  \label{eq:A:U}
  \end{equation}

In regions of space where $\rho(R) = 0$,
the nonlinear equation (\ref{eq:LapU:R})
reduces to a homogeneous equation
with the general solution
  \begin{equation}
  U(R)=k+\frac{m}{R} ,
  \label{eq:Uh}
  \end{equation}
where $k$ and $m$ are integration constants.
Using (\ref{eq:r=UR})--(\ref{eq:A:U}),
to express the line element in the standard form (\ref{eq:ds2:AB}),
one obtains $ B(r) = A(r)^{-1} = k^{-2} \, ( 1 - m/r )^2 $.
If the region of space we are considering extends to $r \to \infty$,
according to the requirements (\ref{eq:AB:inf}),
we set $k=1$, and only $m$ remains as a free parameter.
The line element is then
  \begin{equation}
  \dd s^2 = - \left( 1 - \frac{m}{r} \right)^2 \dd t^2
            + \left( 1 - \frac{m}{r} \right)^{-2} \dd r^2
            + r^2 \, \dd \omega^2,
  \label{eq:ds2:ern}
  \end{equation}
which we recognize as the extremal Reissner--Nordstr{\o}m (ERN) spacetime.
The general Reissner--Nordstr{\o}m (RN) solution
specifies the asymptotically flat spacetime metric
around an electrically charged source.
Expressing the RN line element in the standard form (\ref{eq:ds2:AB}),
one has $ B(r) = A(r)^{-1} = ( 1 - 2m/r + q^2/r^2 ) $,
where $m$ and $q$ are the ADM mass and charge of the source.
For $m<q$, the RN spacetime is regular everywhere except at $r=0$,
while for $m>q>0$, in addition to the singularity at $r=0$,
it exhibits two horizons located at $r = m \pm ( m^2 - q^2 )^{1/2}$.
In the ERN case, i.e. $m=q$
(gravitational attraction balances electrical repulsion),
only one horizon at $r=m$ is present.
It is important to note that the solution (\ref{eq:Uh})
is valid in the range $0 \le R < \infty$ which is according to
eq.\ (\ref{eq:r=UR}) mapped to $m \le r < \infty$.
That is to say that the solution (\ref{eq:Uh})
specifies only the external part of the ERN spacetime
(see~\cite{LeWein04}).

We shall proceed to solve the field equation (\ref{eq:LapU:R})
for several distributions of electrically counterpoised dust
(or extremal charged dust, ECD) $\rho(R)$.
The ECD distributions that will be considered
effectively vanish at large $R$,
so we expect our spacetimes
to behave like (\ref{eq:ds2:ern}) for $R\to\infty$.
The parameter $ m_{\infty} = R\,(U-1)\,|_{R\to\infty}$,
characterizing the asymptotic behaviour of $U(R)$ at large $R$,
is the ADM mass seen by the distant observer.
The behaviour of $U(R)$ as $R \to 0$
can be characterized by the parameter $ m_0 = R\,U\,|_{R\to0}$.
The metric (\ref{eq:ds2:U:R}) is regular at $R=0$
only if $m_0=0$ and, according to (\ref{eq:r=UR}),
the range $0 \le R < \infty$ is mapped to $0 \le r < \infty$.
On the other hand, if $m_0>0$ the function $U(R)$ is infinite at $R=0$
which indicates the singularity.
As the range $0 \le R < \infty$ is now mapped to $m_0 \le r < \infty$,
this singularity is, in the $r$-coordinate, located at $r=m_0$.
Therefore, we may understand the parameter $m_0$
as the `mass below horizon'.
If $\rho=0$ we have $m_0 = m_{\infty}$,
while in case of non-vanishing ECD density $\rho$ we have
  \begin{equation}
  m_{\infty} = m_0 + m_{\rho} ,
  \end{equation}
where $m_{\rho}$ can be understood as the contribution
of ECD to the ADM mass of the configuration.
The `ECD mass' $m_{\rho}$
is the space integral of the ECD density $\rho$:
  \begin{equation}
  \label{eq:mrho}
  m_{\rho} = 4\pi \int_{0}^{\infty}   \rho \,  U^3    \, R^2 \, \dd R
           = 4\pi \int_{m_0}^{\infty} \rho \, A^{1/2} \, r^2 \, \dd r .
  \end{equation}
We conclude this Section by pointing out, for later convenience,
that if certain $U(R)$ and $\rho(R)$ solve the equation (\ref{eq:LapU:R}),
one is allowed to rescale the functions
  \begin{equation}
  \begin{array}{rcl}
  U(R) & \longrightarrow & U^*(R) = U(\alpha R), \\
  \rho(R) & \longrightarrow & \rho^*(R) = \alpha^2 \rho(\alpha R),
  \end{array}
  \label{eq:alpha1}
  \end{equation}
which, as a consequence, rescale the mass parameters according to
  \begin{equation}
  m \longrightarrow m^* = m / \alpha .
  \label{eq:alpha2}
  \end{equation}
These relations will help us compare field configurations
corresponding to different mass parameters.


\section{General ECD distributions and bifurcating solutions \label{sec:bif}}

Our approach in constructing solutions to the Majumdar--Papapetrou systems
is to assume certain ECD distribution $\rho(R)$
and then integrate the differential equation (\ref{eq:LapU:R}) numerically
to obtain the metric function $U(R)$,
and is therefore different from the approach
used in~\cite{BoWick75,LeWein04}
where $\rho(R)$ is analytically reconstructed from the assumed metric
and the requirement that $\rho(R)$ may not be negative.
At the expense of requiring numerical procedures
our approach allows complete freedom in choosing the shape of $\rho(R)$,
which we found more natural.
However, in accord with the requirement
that our solutions be asymptotically flat,
we choose ECD distributions $\rho(R)$
that are well localized in the $R$-coordinate.
More specifically, we require that
the total mass/charge in a linear theory
$4\pi\int_0^{\infty}\rho \, R^2 \, \dd R$ is finite,
i.e.\ that $\rho(R)$ falls off more rapidly than $1/R^3$.
As our first example we take $\rho(R)$ of the form
  \begin{equation}
  \rho(R)=\frac{\eta}{24\pi}\,(R/\tilde{R})\,\ee^{-R/\tilde{R}},
  \label{eq:rho1}
  \end{equation}
where $\tilde{R}$ is the length scale that in further text we set equal to $1$,
$\eta$ has the role of an adjustable source strength factor
(in appropriate units) and $24\pi$ is a normalization constant
that renders $4\pi\int_0^{\infty}\rho\,R^2\,\dd R=1$
for $\eta=1$ and $\tilde{R}=1$.

\begin{figure}
\begin{indented}
\item[] \includegraphics{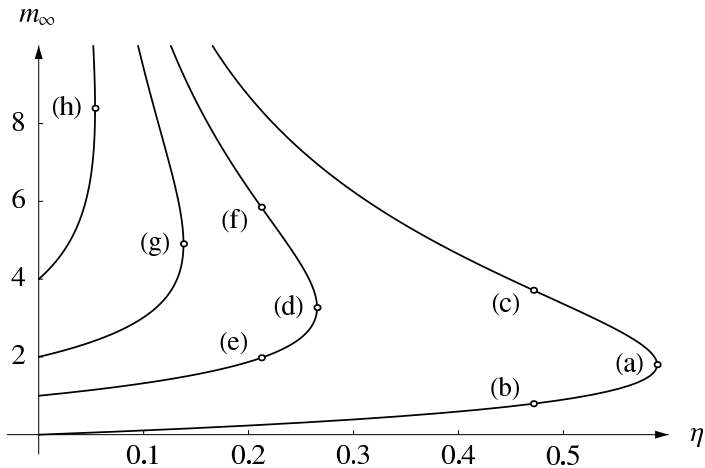}
\end{indented}
\caption{\label{fig:eta}
The dependence of $m_{\infty}$ (ADM mass)
on source strength $\eta$ in solutions to (\ref{eq:LapU:R})
with ECD distribution given by (\ref{eq:rho1})
and boundary conditions $m_0=0$ (leading to regular spacetimes)
and $m_0=1,2,4$ (leading to singular spacetimes).
The numerical values of $\eta$ and $m_{\infty}$
for the solutions labeled by letters $(a)$--$(h)$
are given in table~\ref{tbl:eta}.}
\end{figure}

As (\ref{eq:LapU:R}) is a second order differential equation
we have to impose two boundary conditions onto a solution.
As the first boundary condition,
we use the requirement that the spacetime is asymptotically flat,
i.e.\ $U\,|_{R\to\infty}=1$,
and as the second we chose to fix the parameter $m_0=R\,U\,|_{R\to0}$.
We obtain the solutions
by numerical integration \cite{COLSYS} of the differential equation.
Starting with the solution of (\ref{eq:LapU:R})
for source strength $\eta=0$, i.e.\ $m_\infty=m_0$,
we slowly increase the value of $\eta$.
Increasing of $\eta$ evolves the solution toward $m_{\infty}>m_0$,
but one may only increase the value of $\eta$
up to a critical value $\eta_{\mathrm{c}}$.
The dependence of $m_\infty$ on $\eta$ in solutions obtained in this way
is shown as the lower part of the solution tracks in figure~\ref{fig:eta}.
The critical points are labeled $(a)$, $(d)$, $(g)$, and $(h)$,
and the corresponding values of $\eta_{\mathrm{c}}$ and $m_\infty$
are given in table~\ref{tbl:eta}.

\begin{table}
\caption{\label{tbl:eta}
Numerical values of $\eta$ and $m_{\infty}$
for the solutions labeled by letters in figure~\ref{fig:eta}.}
\begin{indented}
\item[] \begin{tabular}{@{}lllll}
\hline
$m_0$ &
$\eta_{\mathrm{c}}$ &
$m_{\infty}(\eta_{\mathrm{c}})$ &
$m_{\infty}^{-}(\frac{4}{5}\eta_{\mathrm{c}})$ &
$m_{\infty}^{+}(\frac{4}{5}\eta_{\mathrm{c}})$ \\
\hline
0 & $0.589\,272$ & $1.801\,05\;(a)$ & $0.795\,196\;(b)$ & $3.713\,25\;(c)$ \\
1 & $0.265\,212$ & $3.266\,61\;(d)$ & $1.976\,69 \;(e)$ & $5.849\,51\;(f)$ \\
2 & $0.137\,966$ & $4.906\,04\;(g)$ &                   &                  \\
4 & $0.053\,701$ & $8.394\,04\;(h)$ &                   &                  \\
\hline
\end{tabular}
\end{indented}
\end{table}

In addition to these solutions
there exists another class of solutions:
starting from a critical solution,
i.e.~one obtained with $\eta_\mathrm{c}$,
in our numerical procedure,
we could require a slight increase in $m_{\infty}$
and allow the source strength $\eta$ to adjust itself freely.
The solutions to the differential equation can be found
and it turns out that, for this class of solutions,
as we increase $m_{\infty}$, the source strength $\eta$ decreases!
These solutions are indicated as the upper part
of the tracks in the $m_{\infty}$ versus $\eta$ diagram (figure~\ref{fig:eta}).
Therefore, for source strength $\eta>\eta_{\mathrm{c}}$
there appears to be no solution to (\ref{eq:LapU:R})
that is asymptotically flat,
while if starting from $\eta_{\mathrm{c}}$
(i.e., from the critical solution such as
$(a)$ or $(d)$ in figure~\ref{tbl:eta}),
the solutions bifurcate by following either lower
(through points $(b)$ or $(e)$)
or upper $m_{\infty}$-branch (through points $(c)$ or $(f)$).
This kind of bifurcation is of the `turning point type'
as explained in~\cite{Hage} (see also \cite{Sey}).

\begin{figure}
\begin{indented}
\item[] \includegraphics{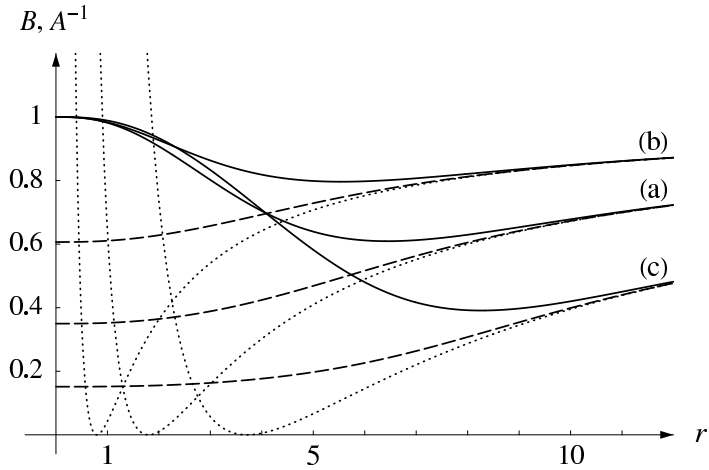}
\end{indented}
\caption{ \label{fig:AB0}
Regular spacetime metric components
$g_{rr}^{-1}=A^{-1}(r)$ (thick lines) and $g_{tt}=B(r)$ (dashed lines)
obtained by solving the field equation (\ref{eq:LapU:R})
with the ECD distribution given by (\ref{eq:rho1})
and boundary condition $m_0=0$.
The solution $(a)$ is the critical solution,
while the solutions $(b)$ and $(c)$ are the two independent solutions
obtained with $\eta=\frac{4}{5}\eta_{\mathrm{c}}$
(see figure~\ref{fig:eta}, and table~\ref{tbl:eta} for numerical values).
The ERN metric components $A^{-1}=B=(1-m_{\infty}/r)^2$
are shown for corresponding $m_{\infty}$ values (dotted lines).}
\end{figure}

In figure~\ref{fig:eta}, the critical point
on the curve corresponding to the boundary condition $m_0=0$
is obtained with $\eta_{\mathrm{c}}=0.589\,27$ and is labeled $(a)$.
The points $(b)$ on the lower and $(c)$ on the upper $m_{\infty}$-branch
indicate the two independent solutions
obtained with $\eta=\frac{4}{5}\eta_{\mathrm{c}}=0.471\,41$.
The corresponding masses are given in table~\ref{tbl:eta}.
The components of the metric in the $r$-coordinate
$A(r)=g_{rr}$ and $B(r)=g_{tt}$
for the solutions labeled $(a)$--$(c)$,
are shown in figure~\ref{fig:AB0}.
At large $r$, the metric components $A(r)$ and $B(r)$
coalesce into the ERN metric (\ref{eq:ds2:ern}).
But in contrast to the ERN metric, $A(r)$ and $B(r)$ of our solutions
are finite at all $r$ so the spacetime does not involve an event horizon.
This is true for all solutions
obtained with the boundary condition $m_0=0$.
We call them the regular solutions,
although in section~\ref{sec:qbh} we show that these
solutions may come arbitrarily close to having an extremal horizon.

\begin{figure}
\begin{indented}
\item[] \includegraphics{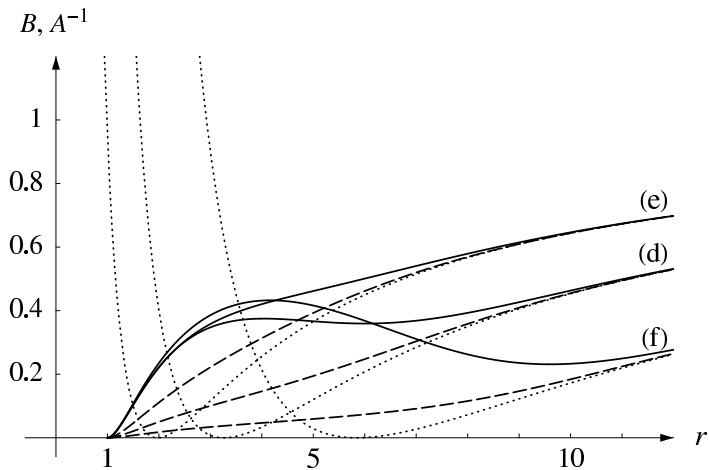}
\end{indented}
\caption{ \label{fig:AB1}
Regular spacetime metric components
$g_{rr}^{-1}=A^{-1}(r)$ (thick lines) and $g_{tt}=B(r)$ (dashed lines)
obtained by solving the field equation (\ref{eq:LapU:R})
with the ECD distribution given by (\ref{eq:rho1})
and boundary condition $m_0=1$.
The solution $(d)$ is the critical solution,
while the solutions $(e)$ and $(f)$ are the two independent solutions
are obtained for $\eta=\frac{4}{5}\eta_{\mathrm{c}}$,
(see figure~\ref{fig:eta}, and table~\ref{tbl:eta} for numerical values).
The ERN profile functions $A^{-1}=B=(1-m_{\infty}/r)^2$
are shown for corresponding $m_{\infty}$ values (dotted lines).}
\end{figure}

The situation is substantially different for the solutions
obtained with the boundary condition $m_0>0$.
The metric components for the case $m_0=1$,
labeled $(d)$--$(f)$, in figure~\ref{fig:eta} and table~\ref{tbl:eta},
are shown in figure~\ref{fig:AB1}.
At $r=m_0$, the metric component $A(r)=g_{rr}$ diverges
($1/A(r)$ reaches zero) which indicates an event horizon.
When approached from $r>m_0$ the functions $1/A(r)$ and $B(r)$
appear to have a double zero at $r=0$ and in this sense this
spacetime singularity is equivalent to the ERN event horizon.
Recall that at $r<m_0$
the spacetime metric is not specified
by the solution to the field equation (\ref{eq:LapU:R}).
This type of solutions we call singular.

For different choices of ECD distribution $\rho$,
as for instance $\rho(R) \propto R^2 \exp(-R^2)$,
basically the same bifurcating behaviour of solutions occurs.
This fact will be used
to simplify the treatment of $\rho$'s in section~\ref{sec:uni}.


\section{The $\delta$-shell ECD distribution \label{sec:delta}}

Here we will consider the case of ECD
distributed on a spherical shell of radius $R_0$.
A situation similar to this one was considered in \cite{Gurses98},
while thick shells were considered in \cite{KZL04}.
We set
  \begin{equation}
  \rho(R) = \eta \, \delta( R - R_0 ) .
  \label{eq:del}
  \end{equation}
Both in the interior ($R<R_0$) and in the exterior ($R>R_0$) space
the general solution to (\ref{eq:LapU:R}) is of the form (\ref{eq:Uh}).
We set the exterior solution to be asymptotically flat
and allow the interior solution to be singular at $R=0$:
  \begin{equation}
  U(R) =
  \left\{ \begin{array}{ll}
    k + m_0/R        \equiv U_{\mathrm{I}}, & R < R_0, \\
    1 + m_{\infty}/R \equiv U_{\mathrm{E}}, & R > R_0.
  \end{array} \right.
  \label{eq:del:U}
  \end{equation}
The requirement that $U_{\mathrm{I}} = U_{\mathrm{E}}$ at $R=R_0$
fixes the value of $k$:
  \begin{equation}
  k = 1 + (m_{\infty} - m_0)/R_0 \, .
  \label{eq:del:k}
  \end{equation}
Integration of the differential equation (\ref{eq:LapU:R})
with the rhs involving the $\delta$-shell source (\ref{eq:del})
yields
  \begin{equation}
  R^2 U' \big|_{R-\epsilon}^{R+\epsilon}
    = - 4\pi \, R_0^2 \, \eta \, U(R_0)^3 \; .
  \end{equation}
When the solution (\ref{eq:del:U}) is substituted
into the above relation the source strength $\eta$, $m_0$,
and the position of the $\delta$-shell source,
are interrelated:
  \begin{equation}
  \eta = \frac{m_{\infty}-m_0}{4\pi\,R_0^2} \,
    \left( 1 + \frac{m_{\infty}}{R_0} \right)^{-3} \; .
  \label{eq:del:eta}
  \end{equation}
This relation leads to bifurcating solutions
similar to those discussed in section~\ref{sec:bif}.

\begin{figure}
\begin{indented}
\item[] \includegraphics{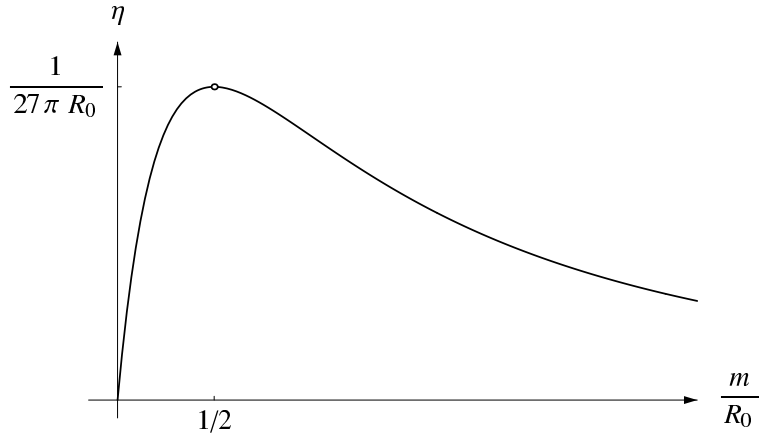}
\end{indented}
\caption{\label{fig:del:eta}
The source strength $\eta$ vs.~$\xi=m/R_0$ in regular solutions
to the field equation (\ref{eq:LapU:R})
with $\delta$-shell source (\ref{eq:del}).}
\end{figure}

\begin{figure}
\begin{indented}
\item[] \includegraphics{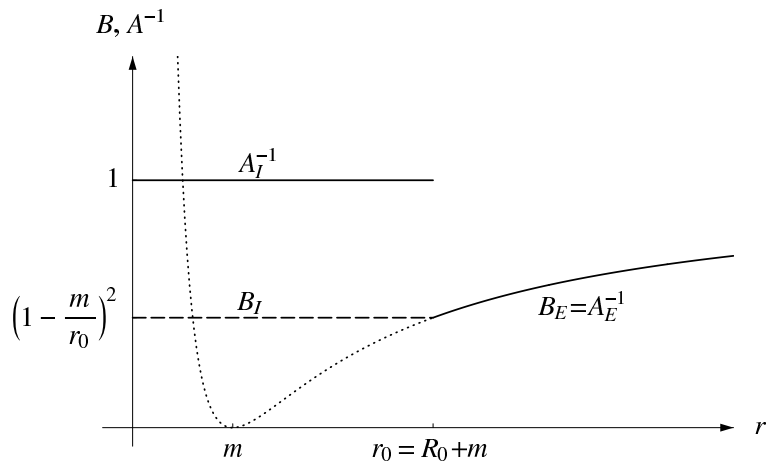}
\end{indented}
\caption{\label{fig:del:AB}
The metric components
$g_{rr}^{-1}=A^{-1}(r)$ (thick line) and $g_{tt}=B(r)$ (dashed line)
for the regular critical solution of the field equation (\ref{eq:LapU:R})
with $\delta$-shell source (\ref{eq:del}).
In the exterior of the shell located at $r_0=m+R_0$,
the metric is equivalent to the ERN (\ref{eq:ds2:ern}) for the mass $m$
(extended by dotted line into the interior space).}
\end{figure}

We now restrict the discussion to the regular solutions,
i.e., we set $m_0=0$
and proceed with only one mass parameter $m=m_{\rho}=m_{\infty}$
given by equation~(\ref{eq:mrho}).
The source strength (\ref{eq:del:eta}) is
  \begin{equation}
  \eta = \frac{1}{4\pi\,R_0} \frac{\xi}{(1+\xi)^3} ,
  \end{equation}
where  $\xi=m/R_0$ and it follows that at $\xi_{\mathrm{c}}=1/2$
there is a maximum $\eta_{\mathrm{c}}=(27\pi\,R_0)^{-1}$
as shown in figure~\ref{fig:del:eta}.
For the regular solution the $R$-coordinate metric component $U(R)$ is
  \begin{equation}
    U(R) = \left\{ \begin{array}{ll}
    1 + m/R_0, & R \le R_0, \\
    1 + m/R,   & R \ge R_0.
    \end{array} \right.
  \end{equation}
According to (\ref{eq:r=UR}), the position of the $\delta$-shell
in the $r$-space is $r_0 = m + R_0$.
In the interior of the $\delta$-shell
the $r$-space metric components are constants
  \begin{equation}
  A^{-1}_{\mathrm{I}}(r) = 1
  \quad \mathrm{and} \quad
  B_{\mathrm{I}}(r) = \left( 1 - \frac{m}{r_0} \right)^2,
  \end{equation}
while in the exterior
the metric follows the ERN metric given by (\ref{eq:ds2:ern}).
The metric components of only the critical regular solution
involving the $\delta$-shell ECD distribution
are shown in figure~\ref{fig:del:AB}.

\section{Bifurcating behaviour
and unified treatment of ECD distributions \label{sec:uni}}

Solutions for different ECD distributions $\rho(R)$
found in sections~\ref{sec:bif} and \ref{sec:delta}
exhibit bifurcating behaviour
with respect to the source strength parameter $\eta$.
When $\eta<\eta_{\mathrm{c}}$, one is able to find
two independent/different solutions
to the same (nonlinear) differential equation
with the same boundary conditions.
As $\eta$ approaches the critical value $\eta_{\mathrm{c}}$
the two solutions become identical.
While the asymptotic flatness
of the two independent solutions is fixed by the boundary conditions,
their ADM mass (\ref{eq:mrho}) is different.

If one takes any $m_{\infty}$ versus $\eta$ curve in figure~\ref{fig:eta}
(and corresponding spacetime metric component solutions),
it is not possible to distinguish the upper branch
from the lower branch solutions
apart from $m_{\infty}$ calculated from (\ref{eq:mrho}).
Only a sequence of solutions provides a means
to determine the branch along which solutions propagate.

One can expect that metric component configurations
leading to smaller ADM mass when a mass source strength $\eta$ is decreased
are physically more acceptable then their counterparts
which produce larger and larger mass $m_{\infty}=m_{\mathrm{ADM}}$.
Indeed, when the source strength is decreased from its critical value
the low-mass bifurcation branch
leads to classical (Newtonian gravity) configurations of ECD.
One may say that the upper branch solutions
(points $(c)$, $(f)$, $(h)$ in figure~1)
may tend to infinite mass which renders them to be (physically) unacceptable.
Further investigation of stability properties of these solutions
\cite{DHSIstab} might give some definite answer to the above assertions.
On the other hand, the spacetimes corresponding
to high-mass configurations have the interesting and important property
of being quasi-singular which we discuss further in section~{\ref{sec:qbh}}.

Based on the bifurcation properties
of the solutions to the field equation (\ref{eq:LapU:R}),
we are going to introduce a normalization of the ECD distributions
appropriate for the Majumdar--Papapetrou (MP) formalism.
This will allow us to treat
the diverse ECD distributions on equal footing
and more easily explore their properties.
We will focus only on the regular solutions,
so only one mass parameter $m=m_{\rho}=m_{\infty}$ will be used.
The ECD distributions considered in section~\ref{sec:bif}
will be used to generate spacetimes that come arbitrarily close
to having extremal horizons.

\begin{figure}
\begin{indented}
\item[] \includegraphics{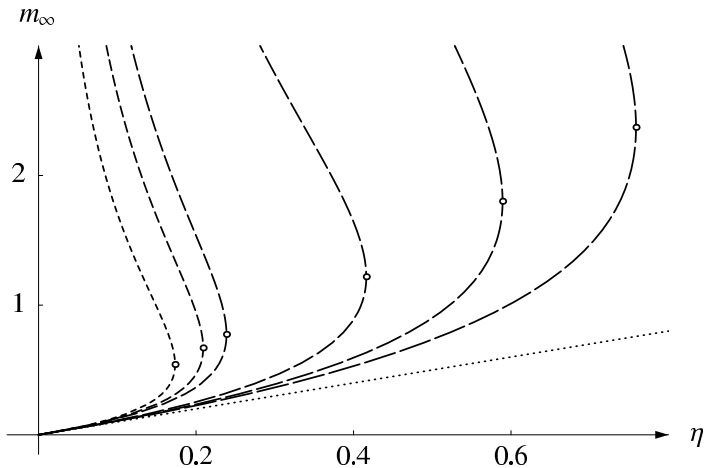}
\end{indented}
\caption{\label{fig:uni:eta}
Dependence of mass $m$ on the source strength $\eta$
for the regular solutions to the field equation (\ref{eq:LapU:R})
with ECD distribution given by (\ref{eq:rho:ab:n}).
Going from left to right, the tracks are ordered
as in the table~\ref{tbl:eta:uni},
critical solutions are indicated by circles.
Dotted line indicates the expected behaviour in the linear theory.}
\end{figure}

We first consider a general distribution $\rho(R) = \eta\,\rho_0(R)$
where $\eta$ is the source strength factor
and $\rho_0$ is normalized so that
  \begin{equation}
  4\pi\int_0^{\infty} R^2\,\rho_0(R)\,\dd R = 1 \;.
  \label{eq:norm:newt}
  \end{equation}
In the linear theory,
the contribution of $\rho = \eta\,\rho_0$
to the total mass and charge of the configuration would be $m=q=\eta$,
and there would be no upper bound imposed onto the source strength $\eta$
(see figure~\ref{fig:uni:eta}, dotted line).
In our case where the field equation (\ref{eq:LapU:R}) is non-linear,
solutions exist only for values of $\eta$
less or equal to a critical value $\eta_{\mathrm{c}}$.
For $\eta=\eta_{\mathrm{c}}$ the solution is unique,
we call it the critical solution,
and we label the corresponding mass with the symbol $m_{\mathrm{c}}$.
For $\eta<\eta_{\mathrm{c}}$ there are two independent, bifurcating solutions,
leading to masses that we label $m^{\pm}$
where it holds $m_{-} < m_{\mathrm{c}} < m_{+}$.

As examples of ECD distributions
normalized according to (\ref{eq:norm:newt}), we considered
  \begin{equation}
  \rho_0(R) = n(a,b)\, \, (R/\tilde{R})^a \, \exp(-(R/\tilde{R})^b) ,
  \label{eq:rho:ab:n}
  \end{equation}
where as before $\tilde{R}=1$ (i.e.\ a unit mass),
$a$, $b>0$ are parameters and $n(a,b) = b/(4\pi\Gamma[(3+a)/b])$.
The dependence of $m$ on $\eta$
for the parameter values $a=0,1,2$ and $b=1,2$
is shown in figure~\ref{fig:uni:eta}.
The numerical values for $\eta_{\mathrm{c}}$ and $m_{\mathrm{c}}$
are given in table~\ref{tbl:eta:uni}.
As can be seen, only the $m^-$-branch,
and only at the low source strength limit,
behaves as it would in the linear theory.

In the context of the field equation (\ref{eq:LapU:R}),
we can formulate a more natural approach to normalization of the sources
for the Majumdar--Papapetrou spacetimes that we will call MP-normalisation.
For the MP-normalized ECD distribution $\hat{\rho}(R)$,
we require that the critical source strength factor
and corresponding mass are both equal to unity,
i.e.\ $\eta_{\mathrm{c}} = m_{\mathrm{c}} = 1$.
To obtain a MP-normalized ECD distribution 
starting from an arbitrarily normalized ECD distribution $\rho_0(R)$,
we first solve (\ref{eq:LapU:R})
to obtain the values $\eta_{\mathrm{c}}$ and $m_{\mathrm{c}}$.
The MP-normalized source $\hat{\rho}$ is then constructed
by rescaling the density $\rho_0$
according to (\ref{eq:alpha1}) and (\ref{eq:alpha2})
with the scaling parameter $\alpha=m_{\mathrm{c}}$.
The MP-normalized source $\hat\rho$ corresponding to $\rho_0$ becomes
  \begin{equation}
  \hat\rho(R) = m_{\mathrm{c}}^2 \, 
    \eta_{\mathrm{c}} \, 
    \rho_0( m_{\mathrm{c}} R ) \; .
  \label{eq:norm:mp}
  \end{equation}
As an example we can take
the $\delta$-shell ECD distribution (\ref{eq:del})
discussed in section~\ref{sec:delta}
for which we obtained $\eta_{\mathrm{c}}=(27\pi R_0)^{-1}$
and $m_{\mathrm{c}}=R_0/2$ (see figure~\ref{fig:del:eta}).
The MP-normalized $\delta$-shell density follows as
$ \hat\rho(R) = (1/54\pi) \, \delta( R - 2 )$.
In the case of the ECD distributions (\ref{eq:rho:ab:n}),
the MP-normalized version reads
  \begin{equation}
  \hat\rho(R) = m_{\mathrm{c}}^{a+2} \, \eta_{\mathrm{c}} \, n(a,b) \,
     R^a \, \exp(-(m_{\mathrm{c}}R)^b) ,
  \label{eq:rho:ab:mp}
  \end{equation}
where $\eta_{\mathrm{c}}$ and $m_{\mathrm{c}}$
obtained numerically are given in table~\ref{tbl:eta:uni}.

The bifurcating solutions found in sections~\ref{sec:bif} and \ref{sec:delta}
have similar properties to the solutions found
long time ago in another nonlinear field theory,
i.e.\ in Yang--Mills gauge theory with external sources \cite{JJR79}.
Stability of solutions has been investigated
and different stability properties
with respect to radial oscillations has been found \cite{Jack80,H86}.
In the context of Einstein--Yang--Mills theory
some of solutions have been found
in the gravitating $SU(2)$ monopole context
\cite{BreitSU2,BrihSU2,LueWein99,Brih00}
and an $SU(3)$ extension of this model \cite{BrihSU3},
which could be characterized as bifurcating.
Similar behaviour has recently been discussed
in an extension of the Standard Model \cite{BrihXSM}.
In some of the above papers the term bifurcation
has been used rather loosely because some solutions
found there do not follow the requirements from literature \cite{Sey,Hage}.
Although an explanation of such behaviour is missing in all those works,
a careful numerical analysis done in \cite{BreitSU2}
as well as \cite{BrihPLB}
offers an opportunity to compare solutions discussed before
and bifurcating behaviour of our solutions.

We end this section by giving some technical details
related to the numerical procedure we used to generate
the bifurcating solutions:
assume a sequence of solutions $f_i(x;\eta_i)$, $i=1,2..c$
to a nonlinear differential equation
obtained with a sequence of source strength factors $\eta_i$.
In order to find an expected upper or lower branch solution,
when a solution is already found,
one can replace the parameter $\eta$ by a function $\eta(x)$
described by differential equation $\dd\eta(x)/\dd x=0$.
Since the system is now enlarged by one first order differential equation,
one additional boundary condition has to be supplied.
This boundary condition
can be constructed from the functional value $f_k(x_s,\eta_k)$,
by requiring that $\widetilde f(x_s,\eta(x_s))=f_k(x_s,\eta_k)\pm\delta f$,
where $\widetilde f(x,\eta(x))$ is a part of the enlarged system,
and $\delta f$ (being a small quantity)
produces upper $(+)$ or lower $(-)$ functional value,
giving an upper or lower branch solution.
It is reasonable to choose $f_k(x_s,\eta_k)$
close to the critical value $\eta_c$.
Once a solution is found
the simple continuation used in the numerical code \cite{COLSYS}
could produce a corresponding sequence of solutions.
Relative tolerances could be chosen to be very stringent
(typical order is $10^{-6}$ to $10^{-8}$)
which assures that bifurcation is not produced by loose numerical boundaries
and justifies the six significant digits given in table~\ref{tbl:eta:uni}.

\begin{table}
\caption{\label{tbl:eta:uni}
The critical points in the regular solutions to (\ref{eq:LapU:R})
obtained with the with the ECD distribution (\ref{eq:rho:ab:n}).}
\begin{indented}
\item[] \begin{tabular}{@{}lllll}
\hline
$a$ & $b$ & $1/n(a,b) $ & $\eta_{\mathrm{c}}$ & $m_{\mathrm{c}}$ \\
\hline
0 & 2 & $\pi^{3/2}$            & $0.173\,340$ & $0.542\,345$ \\
1 & 2 & $2\pi$                 & $0.209\,013$ & $0.670\,181$ \\
2 & 2 & $\frac{3}{2}\pi^{3/2}$ & $0.238\,647$ & $0.774\,735$ \vspace{2pt} \\
0 & 1 & $8\pi$                 & $0.416\,177$ & $1.218\,82$  \\
1 & 1 & $24\pi$                & $0.589\,272$ & $1.801\,05$  \\
2 & 1 & $96\pi$                & $0.758\,753$ & $2.371\,19$  \\
\hline
\end{tabular}
\end{indented}
\end{table}

\begin{figure}
\begin{indented}
\item[] \includegraphics{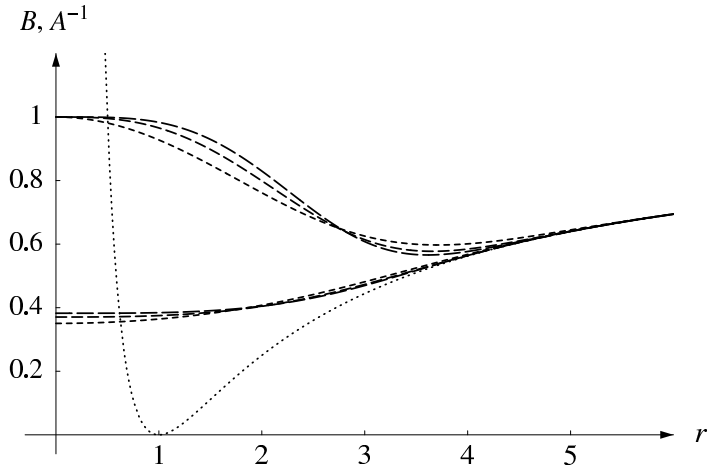}
\end{indented}
\caption{\label{fig:uni:ABc}
The $r$-space metric components $1/A$ and $B$ (dashed lines)
for the critical regular solutions
obtained with the first three MP-normalized sources
of figure~\ref{fig:uni:eta} and table~\ref{tbl:eta:uni} (dashed lines),
and for $m=1$ ERN metric components (dotted line).
(Also to be compared with metric components in figure~\ref{fig:del:AB}.) }
\end{figure}

\section{ECD distributions and quasi black holes \label{sec:qbh}}

\begin{figure}
\begin{indented}
\item[] \includegraphics{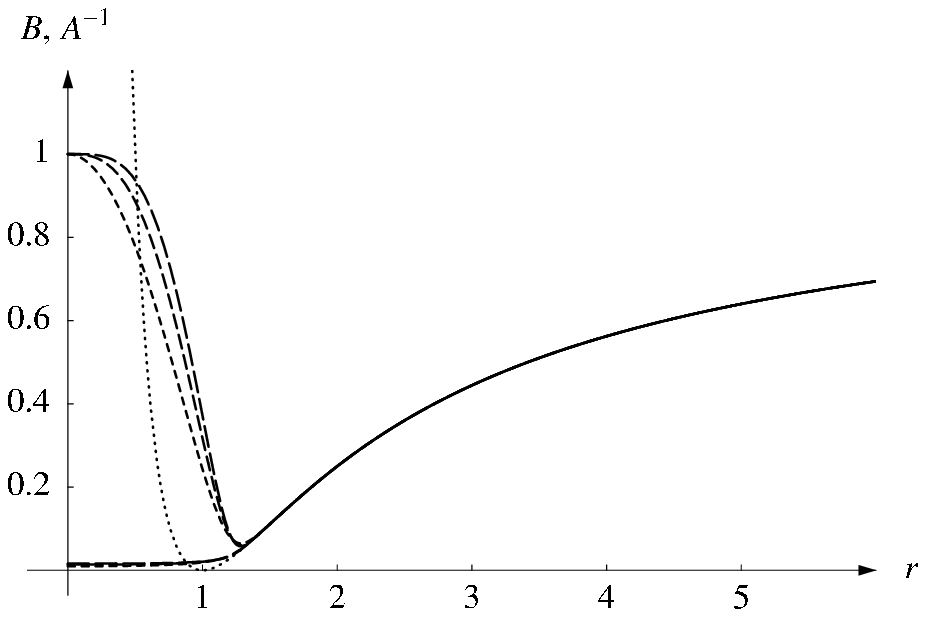}
\item[] \includegraphics{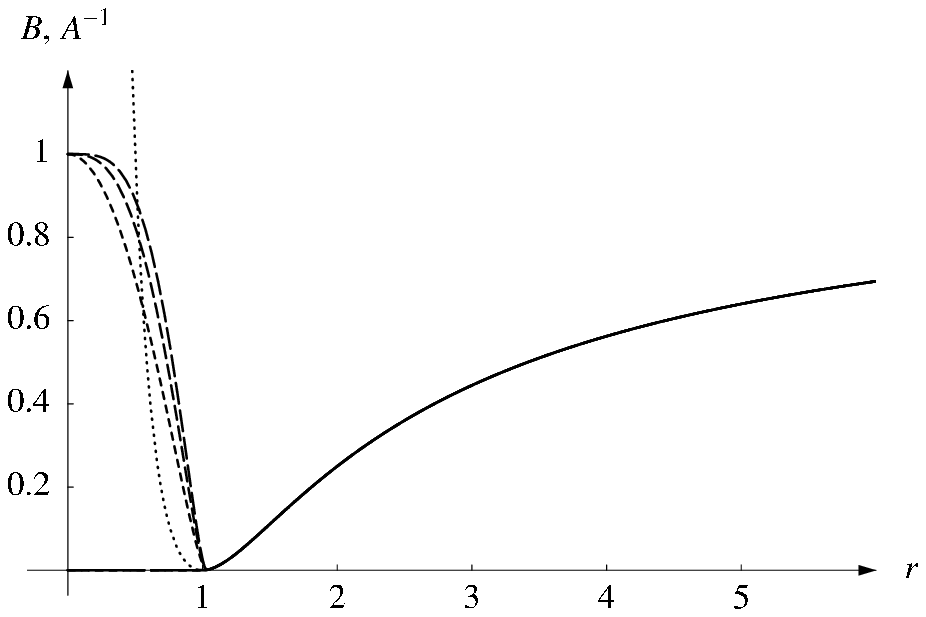}
\end{indented}
\caption{\label{fig:uni:AB10}
The $r$-space metric components
obtained by using the MP-normalized sources as in figure~\ref{fig:uni:ABc}
to generate $m^+=10$ (upper panel) and $m^+=100$ (lower panel)
regular solutions which were then rescaled
to describe unit mass objects.
$m=1$ ERN metric (dotted line). }
\end{figure}

Using a MP-normalized ECD distribution $\hat\rho$,
the critical regular solution to the field equation (\ref{eq:LapU:R})
is obtained with the source strength $\eta=\eta_{\mathrm{c}}=1$.
The $r$-coordinate metric components of the
critical regular solutions to (\ref{eq:LapU:R})
and the ECD distribution (\ref{eq:rho:ab:mp})
with parameters $a=2$ and $b=0,1,2$
are shown in figure~{\ref{fig:uni:ABc}}.
At large $r$, metric components $A$ and $B$
coalesce into the $m=1$ ERN metric,
while at the intermediate values where major part of the ECD is distributed,
they are manifestly regular and do not show significant dependence on the
choice of the shape of the ECD distribution.
By using $\eta<1$, these solutions can,
due to the bifurcating behaviour,
either evolve toward flat space along $m^-$ branch,
or toward higher mass configurations along the $m^+$ branch.
Any of these solutions can be rescaled to describe a unit mass configuration
if the rescaling according to (\ref{eq:alpha1}) and (\ref{eq:alpha2})
is carried out with the scaling parameter $\alpha=m$.
Starting from the critical solutions shown in figure~\ref{fig:uni:ABc}
we followed the $m^+$-branch to obtain the field configurations
corresponding to $m^+=10$ and $m^+=100$
which we then rescaled to restore the unit mass configurations.
The metric components of these solutions are shown in figure~\ref{fig:uni:AB10}.

\begin{figure}
\begin{indented}
\item[] \includegraphics{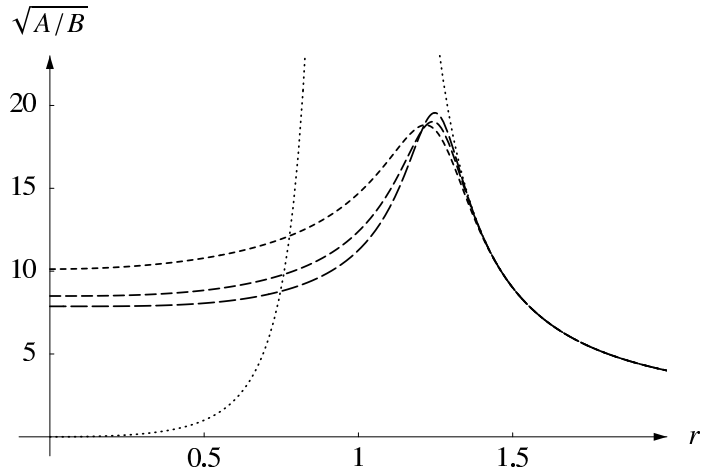}
\end{indented}
\caption{\label{fig:uni:photon}
The ratio $\sqrt{g_{rr}/g_{tt}}=\sqrt{A(r)/B(r)}$
for the solutions shown in figure~\ref{fig:uni:AB10} (dashed lines)
and for the ERN metric (dotted lines).}
\end{figure}

It is plausible that, by following the $m^+$-branch toward higher masses
and then rescaling the solutions to describe the $m=1$ configurations,
the minimum of the function $1/A(r)$ is deeper and closer to $r=1$.
Asymptotically, in the region $r>1$
the metric components $A(r)$ and $B(r)$ follow the ERN metric
and as the point $r=1$ is approached from $r>1$
there appears to be a double zero both in $1/A(r)$ and $B(r)$.
In the $r<1$ region we would asymptotically have $B=0$
which would not allow for timelike intervals.
However, since our spacetimes are regular everywhere
and such a situation is realized only asymptotically,
we can consider a radially moving photon
for which $\dd s^2=0$ and $\dd\omega^2=0$,
so $\dd t / \dd r = \pm \sqrt{A/B}$.
The ratio of the metric components $\sqrt{A/B}$
for the $m^+=10$/rescaled solutions,
related to the time required for the photon
to transverse a unit distance in the $r$-coordinate,
is shown in figure~\ref{fig:uni:photon}.

\begin{figure}
\begin{indented}
\item[] \includegraphics{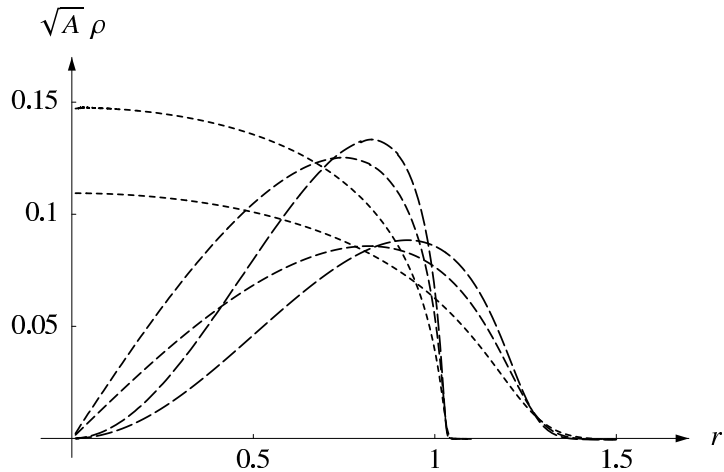}
\end{indented}
\caption{ \label{fig:uni:rho}
The density $\sqrt{A}\rho(r)$ for the solutions in figure~\ref{fig:uni:AB10}.
The densities that effectively vanish at $r\simeq1.1$
correspond to $m^+=100$/rescaled solutions,
while those that effectively vanish at $r\simeq1.5$
correspond to $m^+=10$/rescaled solutions.}
\end{figure}

The distributions of the ECD for the solutions of figure~{\ref{fig:uni:AB10}}
are shown in figure~{\ref{fig:uni:rho}}.
Asymptotically, as we would go higher on the $m^+$-branch
and rescale to restore unit mass configurations,
the density would be completely pulled within the $r=1$ region.
The same was obtained by \cite{LeWein04}.

\section{Conclusions \label{sec:concl}}

The Majumdar--Papapetrou formalism provides a good environment
to study solutions to Einstein--Maxwell equations
where matter is assumed to be described
by electrically counterpoised dust (ECD).
In this paper, we have obtained and analyzed
regular and quasi black hole solutions
stemming from the M--P formalism
and obtained for diverse spherically symmetric ECD distributions
by numerical integration of the nonlinear field equations.
As an immediate consequence of nonlinearity,
bifurcating solutions have been identified
with respect to the amount of ADM mass
allocated in the mass source term.
Also an upper bound to the source strength has been found
above which no solution exists.
Although ECD distributions have assumed
analytically different (spherically symmetric) forms,
we have been able to reformulate the sources
to treat them on equal footing.
From this treatment we have been able to obtain regular solutions
that come arbitrarily close to black hole solutions,
the so called quasi black holes.
Investigation of such bodies could be useful in investigations of
interiors of black holes that still hide many unanswered questions.
Bifurcation is not an unusual feature
in gauge field theory \cite{Jack80} or in gravity \cite{Brih00,LueWein99}.
It is encouraging that here we are able to show
that the bifurcation described in this work
is not an artifact of a particular choice of charge/matter/energy density.
Stability and bifurcation are closely related problems,
so investigation of the above solutions with regard to stability
is a natural extension of this work,
bearing in mind the citation from Ref.~\cite{Sey}
`\dots stability analysis may be more expensive than the
calculation of the solutions themselves.'
%


\section*{Acknowledgments}

This work is supported
by the Croatian Ministry of Science and Technology
through the grant ZP~0036038.
Authors would like to thank for hospitality the
Abdus Salam International Centre for Theoretical Physics,
Trieste, Italy, where part of this work was carried out.




\section*{References}

\begin{thebibliography}{10}

\bibitem{BoWick75}
{Bonnor} W B and {Wickramasuriya} S B P 1975
{\it Mon. Not. R. Astron. Soc.} {\bf 170} 643

\bibitem{BoWick72}
{Bonnor} W B and {Wickramasuriya} S B P 1972
{\it Int. J. Theor. Phys.} {\bf 5} 371

\bibitem{Bonnor98}
{Bonnor} W B 1998
{\it Class. Quantum Grav.} {\bf 15} 351

\bibitem{OppenSny39}
{Oppenheimer} J R and {Snyder} H 1939
{\it Phy. Rev.} {\bf 56} 455

\bibitem{Maj47}
{Majumdar} S D 1947
{\it Phys. Rev.} {\bf 72} 390

\bibitem{Papa47}
{Papapetrou} A 1947
{\it Proc. R. Ir. Acad. Sect. A} {\bf 51} 191

\bibitem{Papa54}
{Papapetrou} A 1954
{\it Z. Phys.} {\bf 139} 518

\bibitem{Varela03}
{Varela} V 2003
{\it Gen. Rel. Grav.} {\bf 35} 1815

\bibitem{Bonnor99}
{Bonnor} W B 1999
{\it Class. Quantum Grav.} {\bf 16} 4125

\bibitem{LeWein04}
{Lemos} J P S and {Weinberg} E J 2004
{\it Phys. Rev. D} {\bf 69}, 104004

\bibitem{KZL04}
{Kleber} A, {Zanchin} V T and {Lemos} J P S 2004
{\em Preprint} gr-qc/0406053

\bibitem{Bonnor60}
{Bonnor} W B 1960
{\it Z. Phys.} {\bf 160} 59

\bibitem{COLSYS}
{Asher} U, {Christiansen} J and {Russel} R D 1981
{\it ACM Trans. Math. Softw.} {\bf 7} 209

\bibitem{Hage}
Hagedorn P 1988 {\it Non-linear oscillations} 2nd.~ed.
(Oxford: Clarendon Press)

\bibitem{Sey}
Seydel R 1988 {\it From Equilibrium to Chaos
-- Practical Bifurcation and Stability Analysis}
(New York: Elsevier)

\bibitem{Gurses98}
{G{\" u}rses} M 1998
{\it Current Topics in Mathematical Cosmology}
(Singapore: World Scientific) p 425

\bibitem{DHSIstab}
{Horvat} D and {Iliji\'c} S submitted to {\it Fizika B (Zagreb)}

\bibitem{JJR79}
{Jackiw} R, {Jacobs} L and {Rebbi} C 1979
{\it Phys. Rev. D} {\bf 20} 474

\bibitem{Jack80}
{Jackiw} R and {Rossi} P 1980
{\it Phys. Rev. D} {\bf 21} 426

\bibitem{H86}
Horvat D 1986 {\it Phys. Rev. D} {\bf 34} 1197

\bibitem{BreitSU2}
{Breitenlohner} P, {Forg\'acs} P and Maison D 1992
{\it Nucl. Phys. B} {\bf 383} 357

\bibitem{BrihSU2}
{Brihaye} Y, {Hartmann} B and {Kunz} J 2000
{\it Phys. Rev. D} {\bf 62} 044088

\bibitem{LueWein99}
{Lue} A and {Weinberg} E J 1999
{\it Phys. Rev. D} {\bf 60}, 084025

\bibitem{Brih00}
{Brihaye} Y, {Grard} F, and {Hoorelbeke} S 2000
{\it Phys. Rev. D} {\bf 62} 044013

\bibitem{BrihSU3}
{Brihaye} Y and {Piette} B M A G 2001
{\it Phys. Rev. D} {\bf 64} 084010

\bibitem{BrihXSM}
Brihaye Y {\it Preprint}  hep-th/0412276

\bibitem{BrihPLB}
{Brihaye} Y, {Hartmann} B and {Kunz} J 1998
{\it Phys. Lett. B.} {\bf 441} 77

\end{thebibliography}
\end{document}